\documentclass[aps,prb,twocolumn,showpacs,longbibliography,superscriptaddress]{revtex4-1}
\usepackage{bm}
\usepackage{graphicx}
\usepackage{epsfig}
\usepackage{hyperref}
\usepackage{natbib}
\usepackage{amsmath}
\usepackage{amssymb}

\newcommand{\rmi}{{\rm i}}
\newcommand{\rmd}{{\rm d}}
\newcommand {\e}{{\rm e}}

\renewcommand {\Re}{\mathop{\mathrm{Re}}\nolimits}
\renewcommand {\Im}{\mathop{\mathrm{Im}}\nolimits}

\newcommand{\init  }{_{\rm i}}
\newcommand{\scat }{_{\rm s}}

\begin{document}

\title{Resonant Brillouin scattering of excitonic polaritons in
multiple-quantum-well structures}
\author{A.\,N.\,Poddubny}
\email{poddubny@coherent.ioffe.ru}
\author{A.\,V.\,Poshakinskiy}
\affiliation{Ioffe Physical-Technical Institute, Russian Academy of
Sciences, 194021 St.~Petersburg, Russia}
\author{B.\,Jusserand}
\affiliation{Institut des Nanosciences de Paris, CNRS UMR 7588, Universit\'e
  Pierre et Marie Curie (UPMC),
F-75005 Paris, France}
\author{A.\, Lema\^\i tre}
\affiliation{Laboratoire de Photonique et de Nanostructures, CNRS, 91460 Marcoussis, France}

\begin{abstract}
We present  theoretical and experimental study of resonant Brillouin scattering of excitonic polaritons in one-dimensional multiple-quantum-well structure. We obtain general analytical results for
the scattering light spectra, valid for arbitrary quantum well  arrangement. Application of our theory to the specific case of short-period
 superlattice shows a perfect quantitative agreement with experimental results for the height, width and position of the  Brillouin scattering peaks and allows us to determine the energy, radiative and nonradiative decay rates of quantum well excitons.  We reveal the signatures of excitonic polariton formation in the scattering spectra and show, that the spectral width and height are strongly sensitive to the number of wells in the sample.
\end{abstract}







\maketitle

\section{Introduction}

Excitonic polaritons play a major role in the optical properties of semiconductors at the energies close to the optically active interband transitions. As demonstrated by Hopfield, excitons are strongly coupled to photons and form mixed excitations propagating in bulk materials, with infinite lifetimes in the absence of defects.\cite{hopfield58}  Resonant optical experiments such as reflectivity and transmission have been widely used to study excitonic polaritons in bulk semiconductors.
The physics of excitonic polaritons has been fully reconsidered in the context of semiconductor multilayers. Both cavity polaritons in photonic microcavities~\cite{kavbamalas} and polaritons in Bragg multiple quantum wells \cite{Ivchenko1994,Ivchenko2005}  have attracted a large amount of work.
 Particularly, considerable progress in theory,\cite{Voronov2004,Poddubny2008prb,Podd_Ivch,Averkiev2012,Poshakinskiy2012} realization, and optical studies 
 \cite{prineas2006,prineas2006apl,Hendrickson2008,Goldberg2009,iorsh2011,chaldyshev2011,chaldyshev2011b,
Chaldyshev2012}
  has been achieved over the last decade, see also the review~[\onlinecite{Ivchenko2013}].

Brillouin scattering of polaritons on acoustic phonons has been demonstrated as a unique tool to directly determine the polariton dispersion in bulk material in the vicinity of the Rabi gap.\cite{Weishuch1982,yu1979} 
The fundamental problem of polariton-mediated inelastic light scattering has been investigated for a long time.\cite{Bendow1970,Bendow1970b,Zeyher1974,Matsushita1984,Matsushita1974b} 
Within the polariton point of view, the inelastic scattering process has been often described as the transformation of a photon at the surface of the solid into a polariton of the same energy, the propagation and scattering of the latter within the solid, and the final propagation and conversion of the scattered polariton into an external photon again at the material boundary. No clear evidence of a specific polariton signature in the resonant Raman cross section could be obtained based on the experimental results on bulk samples available at that time. Bulk excitonic polaritons also present specific theoretical difficulties due to spatial dispersion. 

More recently, polariton scattering by optical phonons in microcavities has been extensively studied and the polariton mediation of light scattering has been clearly evidenced based on the in plane dispersion of cavity polaritons and successfully described using the factorization approach introduced previously.\cite{Fainstein1997,Bruchhausen2008}
This work has been recently extended to acoustic phonons in a double acoustic photonic cavity.\cite{Jusserand2011} 
 Brillouin scattering in multiple quantum wells is likely to display unique properties as compared to what applies to bulk semiconductors.
 Acoustic phonons are folded due to the periodic modulation of the acoustic properties in superlattices.\cite{Jusserand1989}
 Several acoustic modes hence become available for polariton scattering with the energies well above the single Brillouin active acoustic mode in bulk materials. Light scattering intensity by folded phonons in superlattices has been theoretically described based on standard photon scattering in the presence of a modulated photoelastic coefficient.\cite{Jusserand1987} He et al\cite{He1988} later have taken account the modulation of the refractive index between the two materials constituting the superlattice to predict some specific features when the period equals half the scattered wave length. 
 More recently, a form factor has been introduced to describe acoustic light scattering in strong resonance with quantum well excitons\cite{Jusserand2013} and relative intensities of folded acoustic modes have been shown to be governed by their overlap with confined excitons wavefunctions, leading to strong variations with the exciton confinement index.
 Resonant scattering with dispersive exciton polaritons has never been considered theoretically up to now. 
 
Moreover, as the excitons are confined in quantum wells, their inter-well tunneling is strongly suppressed. Contrary to the bulk case, they do not display any dispersion along the superlattice axis. Thus, the dispersion of polaritons in multi-quantum well structure arises only due to the radiative coupling between isolated quantum wells. This system provides a model realization of the ideal Hopfield polariton.

Resonant Brillouin scattering of polaritons in GaAs/AlGaAs multiple quantum wells with the period below the polariton Bragg condition has been recently experimentally investigated.\cite{Jusserand2012} 
Both polariton dispersion and damping have been obtained with a high accuracy. An agreement with a polariton dispersion in a simple Hopfield model in the long wavelength approximation was demonstrated. However, the nontrivial spectra of scattering intensity, as well the effect of the finite structure length on position and width of scattering peaks appears to be beyond that simple model. Explanation of these effects calls for a full theoretical description of the scattering process between incoming and outgoing polariton.
We present in this paper such a theory, and we demonstrate its excellent quantitative agreement with experimental data. We reproduce the dependence of the scattering peak height on the incident light energy and explain its three-peak shape with narrow dips at the exciton resonance frequencies. We also show how the scattering spectra depend on the number of quantum wells in the structure and demonstrate the transition from the single exciton regime to the polariton regime.

The rest of the paper is organized as follows. Sec.~\ref{sec:model} outlines the theoretical approach. Sec.~\ref{sec:results} presents the discussion of the calculated results as well as their comparison to the experimental spectra. Main paper results are summarized in Sec.~\ref{sec:summary}.
\label{sec:intro}


\section{Model}
\label{sec:model}

\begin{figure}[b]
\centering\includegraphics[width=0.8\columnwidth]{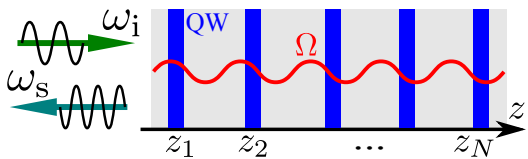}
\caption{Illustration of the Brillouin light scattering in the
multiple-quantum-well structure.  }
\label{fig:simple}
\end{figure}

\begin{figure}[tb!]
\centering\includegraphics[width=\columnwidth]{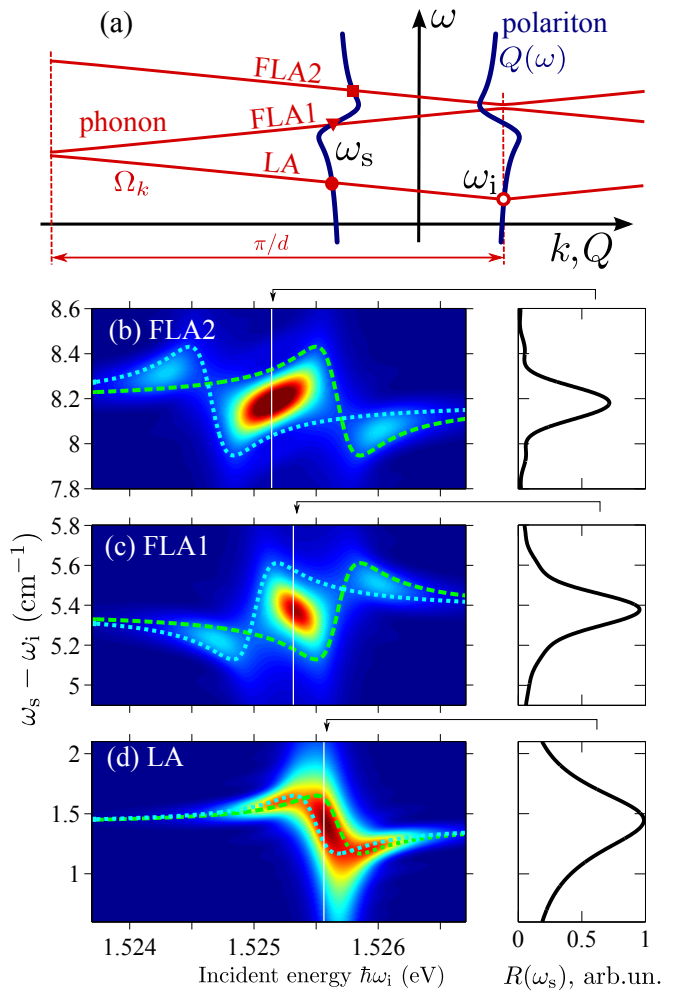}
\caption{(a) Graphical solution of Eq.~\eqref{eq:cases}, corresponding to energy and momentum conservation laws in the anti-Stokes scattering process. For a given incident frequency $\omega\init$ the frequency of scattered light $\omega\scat$ can assume several values corresponding to different branches of folded phonons, LA, FLA1, and FLA2.
Left graphs in panels (b)--(d) show the color map of the anti-Stokes Brillouin scattering intensity as a function of the incident and scattered photon energies. Energy ranges in panels (b),~(c), and (d) correspond to the scattering on FLA2, FLA1 and LA acoustic phonon modes, respectively. Dashed and dotted lines show the contributions of the incident and scattered wave resonances, respectively, to the solution of the conservation laws Eq.~\eqref{eq:cases} and serve as a constraints of the scattering intensity.
Right graphs in panels (b)--(d) present the scattering spectra for the particular values of the incident photon energy $\hbar\omega_{i}$, indicated by vertical white lines in the corresponding map graphs.
 Calculation parameters are indicated in the beginning of Sec.~\ref{sec:results}. Intensity is presented in arbitrary units,
normalization constant is the same for all panels. 
}
\label{fig:map}
\end{figure}

We consider the structure with $N$ quantum wells, embedded in the infinite
matrix with the dielectric constant $\varepsilon_b$, see Fig.~\ref%
{fig:simple}. The dielectric contrast between the wells and the barriers is
neglected. 
Light is normally incident upon the structure at the frequency $\omega\init$. We
are interested in the intensity of the wave at the frequency 
$\omega\scat$, scattered back due to the interaction with phonons. We assume that the interwell spacing is large enough so
that the quantum-mechanical tunneling of the excitons between adjacent wells
can be neglected. The exciton-phonon interaction is introduced by the deformation
potential mechanism.\cite{Cardona}  The polarization $P_j$ of the $j$-th quantum well is described as an
oscillator with the deformation-dependent frequency variation $\epsilon(t)$, 
\begin{equation}  \label{maint}
\frac{dP_j}{dt} + \rmi[\omega_0+\epsilon(t)- \rmi(\Gamma+\Gamma_0)] P_j =
\rmi\xi\Gamma_0 E_j^{\mathrm{(ext)}} \:.
\end{equation}
Here, $\omega_0$ is the exciton resonance frequency, $\Gamma$ and $\Gamma_0$
are exciton nonradiative and radiative decay rates, $\xi= c
n_b/(2\pi\omega_0 a)$ is a dimensionless parameter describing the overlap
between the electric field and the exciton in the quantum well, $a$ is the quantum well width, $E_j^{\mathrm{(ext)}}$ is the external
field that drives the polarization in the $j$-th well and is given by a sum of
the incident wave field $E^{(0)}(t,z_j)$ and the field of the waves emitted
from all other wells, 
\begin{equation}
E_j^{\mathrm{(ext)}}(t) = E^{(0)}(t,z_j) + \frac{\rmi}{\xi}\sum\limits_{l\neq
j} P_l(t-|z_l-z_j|n_b/c) \:.
\end{equation}
The exciton frequency variation $\epsilon(t)$ due to the interaction with
longitudinal acoustic phonons is given by
\begin{equation}  \label{eq:epsilon}
\epsilon(t) = \sum_k (F_k a_k \mathrm{e}^{\rmi kz- \rmi \Omega_k t} + F_k^*
a_k^\dag \mathrm{e}^{-\rmi kz+ \rmi \Omega_k t}) \:.
\end{equation}
Here, $\Omega_k$ is the phonon frequency, $a_k$ and $a_k^\dag$ are the phonon
annihilation and creation operators. The summation runs over the phonon wave
vectors $k$, normal to the wells. In the superlattice, the phonon modes are folded, and the band gaps are formed [Fig.~\ref{fig:map}a]. We are interested in the wave vector regions well separated
 from the superlattice Brillouin zone edges. In this case, we can assume the linear dispersion law $\Omega_k = s |k|$ with $s$
being the speed of sound and the wave vector $k$  defined within the bulk Brillouin zone.
 The exciton-phonon interaction matrix element reads\cite{Jusserand2013} 
\begin{equation}
 F_k \propto \frac{\rmi k }{\sqrt{\Omega_k}} \sum\limits_{\nu=e1,hh1} \Xi_\nu \int |\Phi_\nu (z')|^2 \e^{\rmi k z'} \rmd z'\:,
\end{equation}
where $\Xi_\nu$ is the deformation potential constant for subband $\nu$ and $\Phi_\nu$ is the corresponding wavefunction.

Equations~\eqref{maint}--\eqref{eq:epsilon} completely describe the coupled
system of photons, phonons and excitons. Their solution should be viewed as
an operator acting in the phonon space. Our goal is to find the scattered
light intensity at the frequency $\omega\scat$, averaged over the phonon
distribution. To this end, we start with a Fourier
transformation of Eq.~\eqref{maint} with respect to the time $t$. The result
in the frequency domain assumes the form 
\begin{multline}  \label{main}
(\omega_0 - \omega - \rmi \Gamma) P_j(\omega) - \rmi \Gamma_0 \sum_{l=1}^N 
\mathrm{e}^{\rmi q(\omega) |z_j-z_l|} P_l(\omega) \\
= \xi\Gamma_0E_j^{(0)}(\omega) - V_j(\omega) \:,
\end{multline}
where $q(\omega) =\omega n_b/c$ is the light wave vector corresponding to
the frequency $\omega$ and the potential $V_j$ describes the exciton-phonon
interaction, 
\begin{multline}  \label{Vomega}
V_j(\omega) = \sum_k \bigl[F_k a_k \mathrm{e}^{\rmi kz_j } P_j(\omega - \Omega)
\\
+F_k^* a_k^\dag \mathrm{e}^{-\rmi kz_j } P_j(\omega + \Omega)\bigr] \:.
\end{multline}

In order to find the scattered light intensity we solve Eq.~\eqref{main}
 assuming that the exciton-phonon interaction is weak and using the perturbation theory.
The Fourier component of the electric field of the incident monochromatic wave with the frequency $\omega\init$ reads $E^{(0)}_j(\omega)=E\init {\rm e}^{\rmi q(\omega\init ) z_j}
\delta(\omega-\omega\init )$. The incident electric field induces the polarization in the $j$-th quantum well  
\begin{equation}  \label{eq:P0}
P_j^{(0)} (\omega)=\xi\sum\limits_{l=1}^N G_{jl}(\omega) E^{(0)}_l(\omega)\:,
\end{equation}
where $G_{jl}(\omega)$ is the Green function of the unperturbed Eq.~%
\eqref{main}, defined from the matrix equation\cite{Ivchenko2005,Voronov2007}
\begin{equation}  \label{eq:defG}
(\omega_0-\omega-\rmi\Gamma)G_{jl}(\omega)-\rmi\Gamma_{0}\sum_{m=1}^N\mathrm{e}%
^{\rmi q(\omega)|z_j-z_m|}G_{ml}(\omega)=\Gamma_0\delta_{jl}\:.
\end{equation}
Polarization Eq.~\eqref{eq:P0} oscillates at the same frequency as the initial wave and
describes coherent response, such as reflection and transmission.
Interaction with the acoustic phonons induces the polarization at the
shifted frequency $\omega\scat \ne \omega\init$,
\begin{equation}  \label{eq:P1}
P_j^{(1)} (\omega\scat ) = \sum\limits_{l=1}^N G_{jl}(\omega\scat )
V_l^{(1)}(\omega\scat ) \:,
\end{equation}
where $V_l^{(1)}(\omega\scat )$ is given by Eq.~\eqref{Vomega} with $%
P_j(\omega)$ replaced by $P_j^{(0)}(\omega)$ from Eq.~\eqref{eq:P0}. The
phonon-induced polarization Eq.~\eqref{eq:P1} radiates into the scattered
wave 
\begin{equation}  \label{eq:Es}
E\scat (\omega\scat ) = \frac{\rmi}{\xi} \sum\limits_{j=1}^N P_j^{(1)}(\omega%
\scat ) \mathrm{e}^{iq(\omega\scat )z_j}\:.
\end{equation}
The scattered light spectrum is given by 
\begin{equation}  \label{eq:Iscat}
I\scat (\omega) = \int_{-\infty}^{\infty} \langle E\scat ^\dag(0) E\scat %
(t)\rangle \mathrm{e}^{\rmi\omega t} dt\:,
\end{equation}
where the angular brackets denote the averaging over the phonon distribution
and $E\scat(t)=\int_{-\infty}^\infty E\scat (\omega)\mathrm{e}^{-\rmi\omega
t}d\omega/(2\pi)$. Substituting Eqs.~\eqref{eq:P0}--\eqref{eq:Es} into Eq.~%
\eqref{eq:Iscat} we get $I\scat (\omega\scat )= R(\omega\scat  , \omega\init
) |E\init  |^2$, where 
\begin{multline}
R(\omega\scat , \omega\init ) = \sum_k |F_k|^2 \bigl[ \bar n_k \left|
S(\omega\scat , \omega\init ; k) \right|^2 \delta(\omega\scat - \Omega_k-\omega%
\init ) \\
+ (\bar n_k+1) \left| S(\omega\scat , \omega\init, -k) \right|^2
\delta(\omega\scat + \Omega_k-\omega\init )\bigr]\:,  \label{eq:R1}
\end{multline}
$\bar n_k=\langle a_k^\dag a_k\rangle$ is the average number of phonons with
wave vector $k$ and $S(\omega\scat , \omega\init  ; k)$ is the scattering
factor, 
\begin{align}
S(\omega\scat , \omega\init ; k) &= \sum_{j=1}^Ng_{j}(\omega\init %
)g_{j}(\omega\scat )\mathrm{e}^{\rmi k z_j}\:, \\
g_j(\omega)&=\sum\limits_{m=1}^N\mathrm{e}^{\rmi q(\omega) z_m}G_{jm}(\omega)
\label{eq:gm} \\
&=-\rmi[G_{1j}(\omega)(\omega_0-\omega-\rmi\Gamma)/\Gamma_0-\delta_{1j }]
\:.  \notag
\end{align}
We used here the reciprocity of the Green function, $G_{ij}(\omega) = G_{ji}(\omega)$, and  the Green function definition Eq.~\eqref{eq:defG} with $j=1$.

Equation~\eqref{eq:R1} describes the scattered light spectra in the general case of arbitrary positions of the quantum wells. Below we use this result for the case when the Green function can be written out in an explicit form. We consider the periodic structure with the period $d$, $z_j=(j-1)d$. Then the Green function has the form\cite{Voronov2007} 
\begin{align}\label{eq:Gper}
&G_{mn} =  \chi \delta_{mn} + P \mathrm{e}^{\rmi Q |z_m-z_n|}+\frac{r_\infty P}{1-r_\infty^2\mathrm{e}^{2\rmi QL}} \\\nonumber
\times& \Bigl[ \mathrm{e}^{\rmi Q(z_m+z_n)} + \mathrm{e}^{\rmi Q(2L-z_m-z_n)}+
2r_\infty\mathrm{e}^{2\rmi QL} \cos (z_m-z_n) \Bigr] ,  
\end{align}
where $r_\infty=-[1-\mathrm{e}^{-\rmi(q-Q)d}]/[1-\mathrm{e}^{-\rmi(q+Q)d}]$
is the light reflection coefficient from the infinite structure, $L=d(N-1)$
is the structure thickness, $\chi = \Gamma_0/(\omega_0-\omega-\rmi\Gamma)$, $ P= \rmi\chi^2 \sin qd/\sin Qd$, and $Q$ is the polariton wave vector defined by\cite{Ivchenko1991} 
\begin{equation}
\cos Qd = \cos qd -\chi \sin qd \:.  \label{eq:Q}
\end{equation}
This simplifies the expression for the structure factor Eq.~%
\eqref{eq:gm} to 
\begin{equation}
g_j=-\rmi\frac{r_1(1-r_\infty)(\mathrm{e}^{\rmi Q z_j}+r_\infty\mathrm{e}^{\rmi Q
(2L-z_j)})}{1-r_\infty^2\mathrm{e}^{2\rmi QL}} \:,
\end{equation}
where $r_1 = \rmi\Gamma_0/[\omega_0-\omega-\rmi(\Gamma_0+\Gamma)]$ is the single
quantum well reflection coefficient.

In the case of short-period superlattice, $|q(\omega_0)d|\ll 1$, relevant to the experimental data discussed in Sec.~\ref{sec:results}, the polariton dispersion relation Eq.~\eqref{eq:Q} can be simplified and reduces to the effective medium approximation,\cite{Ivchenko2005} 
\begin{equation}
Q=\frac{\omega}{c}\sqrt{\varepsilon_{\mathrm{eff}}(\omega)},\quad
\varepsilon_{\mathrm{eff}}= n_b^2\left( 1+\frac{\omega_{\mathrm{%
LT}}}{\omega_0-\omega-\rmi\Gamma}\right)\:,  \label{eq:Q_eff}
\end{equation}
where the effective longitudinal-transverse splitting is given by $\omega_{\mathrm{LT}}=2\Gamma_0/\sin q(\omega_0)d$.

Using the Green function Eq.~\eqref{eq:Gper} we arrive to the final expression for the scattered light intensity, 
\begin{align}
&R(\omega\scat ,\omega\init ) = Y_kA(\omega\init )A(\omega\scat )\label{eq:Rfinal}\\\nonumber
&\times \left[\left| B\left( \omega\init,\omega\scat ,\frac{|\omega\init-\omega\scat|}{s}\right)\right|^{2} + \left|B\left(\omega\init ,\omega\scat ,-\frac{|\omega\init-\omega\scat|}{s}\right)\right|^{2}\right] ,  
\end{align}
where 
\begin{align} 
 \label{eq:B}
Y_k=\frac{\bar n_k|F_{k}^2|}{2\pi s\Gamma_0^2}\:,\quad A(\omega)&=\left|\frac{%
r_1(1-r_\infty)}{1-r_\infty^2\mathrm{e}^{2\rmi QL}}\right|^2,   
\\\label{eq:B}
B(\omega\init ,\omega\scat ,k)=&\Delta(k+Q\init +Q\scat)   \\
+ r_{\infty}(\omega\init)\,\mathrm{e}^{2\rmi Q\init L}&\Delta(k-Q\init +Q\scat )   \notag
\\
+ r_{\infty}(\omega\scat)\,\mathrm{e}^{2\rmi Q\scat L}&\Delta(k+Q\init -Q\scat )   \notag
\\
+ r_{\infty}(\omega\scat)r_{\infty}(\omega\init)\,\mathrm{e}^{2\rmi (Q\init +Q\scat )L}&\Delta(k-Q\init %
-Q\scat )\:,    \notag \\
\Delta(Q)=\sum\limits_{m=1}^N\mathrm{e}^{\rmi Qz_m}&= \frac{1-\mathrm{e}^{\rmi
QdN}}{1-\mathrm{e}^{\rmi Qd}},    \label{eq:Delta}
\end{align}
and $Q\init=Q(\omega\init)$, $Q\scat=Q(\omega\scat)$.
Eq.~\eqref{eq:Rfinal} has been
derived assuming that the phonon energy is much less than the
temperature $T$, so that $\bar n_k\sim T/(\hbar\Omega_k)\gg 1$.
This assumption is valid for experimental conditions of Ref.~%
[\onlinecite{Jusserand2012}], performed at $T=80$~K with typical phonon
energies less than 1\,meV. 

The structure of Eq.~\eqref{eq:Rfinal} can be understood as follows. The
factors $A(\omega)$ describe the propagation of incident and scattered waves
inside the system. The quantity $B(\omega\init  ,\omega\scat,k)$ in Eq.~%
\eqref{eq:B} is the structure factor for the scattering of polaritons with
emission/absorption of the phonon with the wave vector $k$. First term in
Eq.~\eqref{eq:B} describes direct scattering, when the scattered polaritonic
wave propagates oppositely to the incident one. Remaining three terms
correspond to the scattering on the phonon combined with the polariton
reflection from the structure boundary. In case of the infinite structure
these three terms vanish while the quantity $|\Delta(Q)|^2$ reduces to the $%
\delta$-function, reflecting the (quasi)momentum conservation law, 
\begin{equation}
\lim_{N\to\infty}|\Delta(Q)|^2=2\pi\sum\limits_{m=-\infty}^\infty\delta(Qd-2\pi
m)\:.  \label{eq:Delta2}
\end{equation}
In the structure with finite number of wells $N$ the momentum conservation
law is relaxed, and the $\delta$-function in Eq.~\eqref{eq:Delta2} is replaced by the peak of finite width of the order $\pi/(Nd)$.


\section{Results and discussion}\label{sec:results}

In this section we apply our general theory of Brillouin light scattering in quantum well structures to the case of short-period sample with the period  $d=24.6$\,nm and $N=40$ wells, studied before in Ref.~[\onlinecite{Jusserand2012}]. In the calculations we use the parameters $s=5\times 10^5$\,cm/sec, $\varepsilon_b=11.8$, $\hbar\Gamma=0.17$\,meV, $\hbar\Gamma_0=39\,\mu$eV, $\hbar\omega_0=1.52565$\,eV, that provide the best fit of the experimental data presented in Sec.~\ref{sec:exp}.

\subsection{Overview of the scattering map}\label{sec:gen}

From the general expression for the scattered light intensity Eqs.~\eqref{eq:Rfinal}--\eqref{eq:Delta} it follows that for a given incident light frequency $\omega\init$ 
 the maximum of the  scattered light spectrum
in long enough structures should correspond to the  frequency $\omega\scat$ that fulfills the conservation law for momentum, $\Re Q\init+\Re Q\scat \pm k =2p\pi/d$, where $k$ is the phonon wave vector fixed by the energy conservation law $|\omega\scat-\omega\init|=\Omega_k=s|k|$, and $p$ is an arbitrary integer that specifies the branch of folded phonons. Figure~\ref{fig:map}(a) shows a graphical solution of the equations corresponding to the conservation laws.
We are interested in  small values of the parameter $p$, where the scattering is more intense. We consider three cases, marked as LA, FLA1 and FLA2; the conservation laws then assume the form
\begin{equation}\label{eq:cases}
|\omega\scat-\omega\init|=s\times%
\begin{cases}
\Re (Q\scat+Q\init)\,, & \text{(LA)}, \\ 
2\pi /d-\Re (Q\scat+Q\init)\,, & \text{(FLA1)}, \\ 
2\pi /d+\Re (Q\scat+Q\init)\,, & \text{(FLA2)}. 
\end{cases}%
\end{equation}
Solving Eq.~\eqref{eq:cases} one can determine
the frequency $\omega\scat$ and, hence, the Brillouin shift $|\omega\scat-\omega\init|$  for each incident frequency $\omega\init$.

Figure~\ref{fig:map} shows the color map of the dependence of the anti-Stokes  scattering intensity  on the incident and scattered photon energies. The incident photon energy is chosen to be close to the exciton resonance frequency $\omega_0$ while that of the scattered photon is varied near the frequencies corresponding to  scattering on LA, FLA1 and FLA2 phonon modes, see panels (d),~(c), and~(b) in Fig.~\ref{fig:map}, respectively.

In order to find the frequency $\omega\scat$ that fulfills the conservation laws, and thus realizes the maximum of the scattered light spectra, we solve Eq.~\eqref{eq:cases} iteratively. First we neglect the exciton contribution to the polariton dispersion assuming $Q(\omega)=n_b \omega/c$.
This allows us to obtain the value $\bar\Omega$ of the bulk Brillouin shift $|\omega\scat-\omega\init|$ that is valid far from the exciton resonance,
\begin{equation}
\bar{\Omega}=2s\times
\begin{cases}
\omega _{0}n_{b}/c\,, & \text{(LA)}, \\ 
\pi /d-\omega _{0}n_{b}/c\,, & \text{(FLA1)}, \\ 
\pi /d+\omega _{0}n_{b}/c\,, & \text{(FLA2)}.%
\end{cases}
\label{eq:barOmega}
\end{equation}

Equation~\eqref{eq:barOmega} describes the exciton-independent contribution to the Brillouin shift. 
In Fig.~\ref{fig:map}(a)--(c) we plot  two approximations to the Brillouin shift, that take exciton into account. The first one  corresponds to the terms $Q\init+Q\scat$ in Eq.~\eqref{eq:cases} estimated as $2Q\init$ and is shown by the dashed curves. The second one, shown by the dotted curves, corresponds to $Q\init+Q\scat$ replaced by $2Q\scat=2Q(\omega\init+\bar \Omega)$ and can be obtained shifting the first one  by the frequency $\bar\Omega$ along the horizontal axis.
These two approximations serve as constraints of the scattering intensity. The most intense scattering takes place in the areas of the map bounded by these two curves.  For LA case the bulk Brillouin shift $\bar\Omega$ is smaller than the exciton resonance width and these curves nearly coincide, so the dependence of the maximum position on the initial wave frequency exhibits one resonance. For the FLA1 and FLA2 scattering the shift $\bar\Omega$ is significant and two resonances, the incident one and the scattered one, can be clearly distinguished.

Position of the scattering spectra maximum for given incident frequency $\omega\init$ is close to the half-sum of the two curves, i.e. it can be well approximated substituting $Q\scat=Q(\omega\init \pm \bar\Omega)$ into the expression for the Brillouin shift Eq.~\eqref{eq:cases}; here plus and minus correspond to anti-Stokes and Stokes processes, respectively, and the polariton wave vector is determined from Eqs.~\eqref{eq:Q},\eqref{eq:Q_eff}.

\subsection{Comparison to experimental data}\label{sec:exp}

\begin{figure*}[t]
\includegraphics[height=0.5\textwidth]{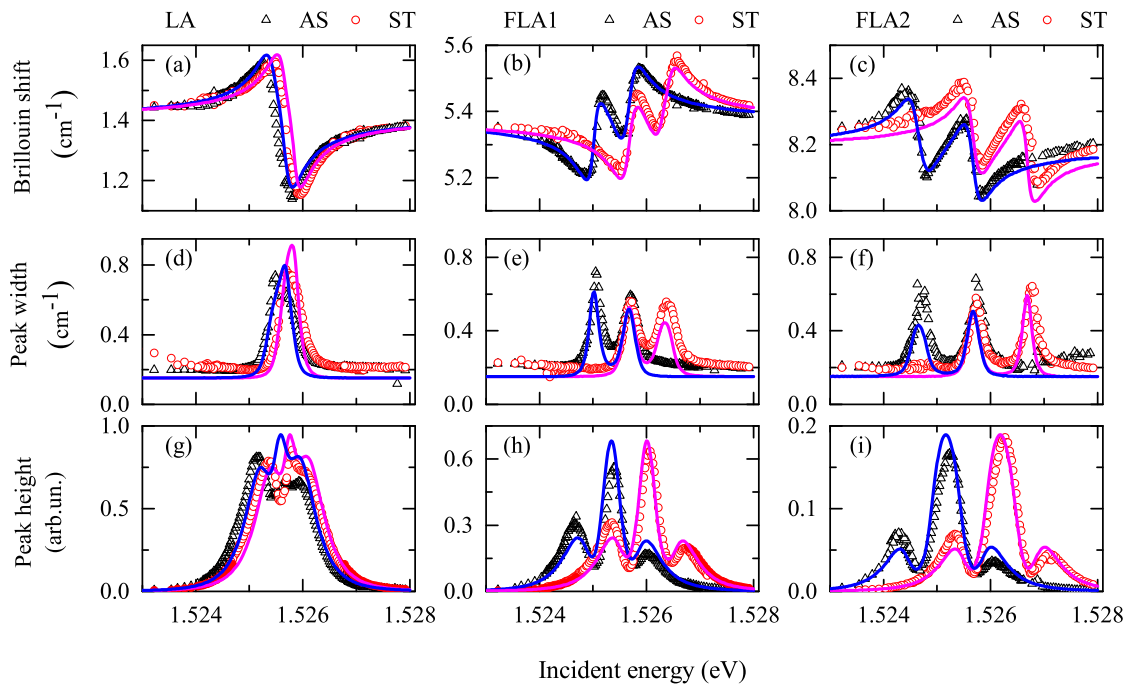}
\caption{Dependence of the scattering peak position (a)--(c), width (d)--(f) and height
(g)--(i) on the incident photon energy. Scattering on the LA, FLA1, and FLA2 phonon modes corresponds to panels (a,d,g), (b,e,h), and (c,f,i), respectively. Points show the experimental data while lines present the theoretical fit with the parameters indicated in the beginning of Sec.~\ref{sec:results}. Magenta lines/red circles and blue lines/red triangles correspond to Stokes and anti-Stokes scattering,
respectively. Vertical scale for panels (g)--(i) is the same.}
\label{fig:fit}
\end{figure*}

For each value of the incident photon energy $\hbar\omega\init  $
the Brillouin scattering intensity  as a function of
the scattered energy $\hbar\omega\scat $ has a single peak, see right panels in Figs.~\ref{fig:map}(b)--(d). This peak is characterized by its position (i.e., Brillouin shift), full width at half-maximum, and its height. Figure~\ref{fig:fit} presents the
comparison of the experimental (points) and fitted theoretical
 (curves) dependences of these three parameters on the incident energy $\hbar\omega%
\init $.  The best fit has been obtained for the parameters indicated in the beginning of Sec.~\ref{sec:results}.
In the fit we neglected the dependence of the factor $Y_k$ on $k$ within a given phonon branch. In order to describe experimental peaks peak height spectra the ratio of the factors $Y_k$ between the phonon branches was taken to be $Y_{\rm LA}$\,$:$\,$Y_{\rm FLA1}$\,$:$\,$Y_{\rm FLA2} = 1$\,$:$\,$0.8$\,$:$\,$0.3$. This agrees well with the approximation of rectangular quantum well with infinitely high barriers and the width equal to $17$\,nm that yields the ratio $1$\,$:$\,$0.7$\,$:$\,$0.4$.

The presented data are slightly different from those published in Ref.~\onlinecite{Jusserand2012} due to the improvement of the temperature stabilization at 70~K and of the resolution of the spectrometer. Raman spectra have been obtained with a high resolution Jobin Yvon U1000 double spectrometer equipped with a low noise CCD multichannel detector. As emphasized in Ref.~\onlinecite{Jusserand2012}, the resonance is so strong that both Brillouin lines and the Rayleigh line at the laser frequency could be measured without spectral filtering and using the same time of integration. As in Ref.~\onlinecite{Jusserand2012}, we used a Coherent MBR 110-PS single line Ti-Sapphire tunable laser with a negligible linewidth below 5~MHz and stability of the order of 0.01 cm$^{-1}$ over long periods.  With an entrance slit of 20~$\mu$m, the laser line had a full width at half maximum of 0.11 cm$^{-1}$ giving the resolution of the system. The sample was mounted on the cold finger of a Helium circulation cryostat with a temperature stability of around 10$^{-2}$ K over hours of experiment. This stability appeared to be critical to avoid fluctuations of the exciton energy during the full measurement of the resonance curve. The incident power was kept as low as possible and measurements have been done with 3 $\mu$W focalized with a 14 cm focus lens.  

Figures~\ref{fig:fit}(a)--(c) show dependence on the incident frequency of the Brillouin shift, deduced as a position of the maximum in the scattered light spectrum. In agreement with the above theory it has two resonances, corresponding to the incident and the scattered polaritons, that are split by the value bulk Brillouin shift $\bar \Omega$ given by Eq.~\eqref{eq:barOmega}. This resonances merge in the LA case and are well-resolved in FLA1 and FLA2 cases. One can see a perfect agreement of theory and experiment.

The dependence on the incident frequency of the scattering peak width is shown in Figs.~\ref{fig:fit}(d)--(f). The spectrum has two peaks, corresponding to the resonances in the Brillouin shift
spectra.  The peak width values are due to the polariton absorption and can be roughly estimated as $4s\Im Q(\omega_0)\approx 0.7~\mathrm{cm^{-1}}$.\cite{Jusserand2012,brenig1972} Experimental peak width is
systematically larger than the calculated one  that can be explained by the finite resolution of the
experimental setup. Importantly, both experimental and theoretical values of
the width do not tend to zero at large detuning from the resonance, but
reach a plateau instead. This plateau is the effect of the finite size of
the structure. Corresponding background value of the width is approximately given by $2s\delta Q\approx 0.16~\mathrm{cm^{-1}}$, where $\delta Q=\pi/(Nd)$ is the wave vector uncertainty due to
the finite size of the system. The explanation of this plateau presents an advantage of
the current approach over the model of an infinite structure used in Ref.~\onlinecite{Jusserand2012}.
A more detailed discussion of the peak width spectra for structures with different number of quantum wells is given in Sec.~\ref{sec:finite}.

Figures~\ref{fig:fit}(g)--(i) present the dependence of the scattering peak height on the incident frequency. The peak height spectra anticorrelate with the peak width spectra, cf.~panels (g)--(i) and (d)--(f). This is because the scattering peak height is at maximum when the polariton absorption is small, which corresponds to the minima of peak width spectra. 
As a result, there are three peak height  maxima located around the two resonances for initial and scattered waves. These maxima are hardly distinguishable for LA mode and well resolved for the FLA1 and FLA2 modes.
The three-peak structure of the peak height spectra is beyond the simplified approach of Ref.~[\onlinecite{Jusserand2012}] and, as shown is Sec.~\ref{sec:finite}, originates from the polariton formation.

\subsection{Dependence on damping and finite size effects}\label{sec:finite}

\begin{figure}[tb!]
\centering\includegraphics[height=0.7\columnwidth]{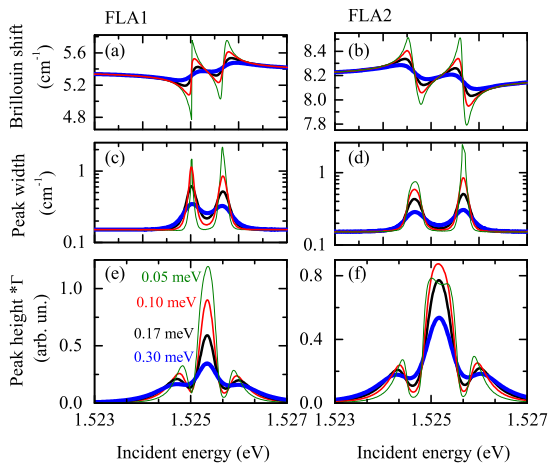}
\caption{Dependence of the Brillouin shift (a,b), scattering peak width (c,d) and height (e,f)
on the incident photon energy for anti-Stokes scattering on the FLA1 (a,c,e) and FLA2 (b,d,f) phonon branches. 
Curves are plotted for different values of the exciton nonradiative damping $\Gamma$ indicated in the graph.
The thickness of the line increases with the value of $\Gamma$,
the black curves of intermediate thickness correspond to the value $\hbar\Gamma=0.17$~meV obtained from the fit of experimental data. Other parameters are indicated in the beginning of Sec.~\ref{sec:results}. For better presentation the peak height spectra have been multiplied by the corresponding values of $\Gamma$, the scale in panels (e) and (f) is the same.}
\label{fig:Gamma}
\end{figure}

In this section we discuss the physical meaning of the observed resonant behavior with the help of a further analysis of the effects of exciton non-radiative damping $\Gamma$ and of a finite number of wells $N$ on the calculated quantities. Nonradiative damping is a key mechanism limiting the formation of polaritons and governing the amplitude and the width of the peculiarities at the exciton resonance energy in the real and  imaginary parts of the polariton dispersion.  We describe the experimental data with a remarkably small damping parameter, $\hbar\Gamma = 0.17$\,meV. Our model allows us to investigate how critical this value is for the determination of the quantities relevant to the polariton scattering. 
Figure~\ref{fig:Gamma} shows the Brillouin shift (a,b), scattering peak width (c,d) and height (e,f) of FLA1 (a,c,e) and FLA2 (b,d,f) anti-Stokes scattering for four different values of $\Gamma$. As expected, smaller values provide more structured spectra: the oscillations of the Brillouin shift become steeper, the peaks in the linewidth become higher and narrower while the dips in the intensity become deeper.  For the smallest value of $\hbar\Gamma=0.05 $~meV $\sim \hbar\Gamma_{0}$, an additional dip appears in the central maximum.
 On the other hand, for $\hbar\Gamma = 0.3$\,meV, less than  twice the value relevant to our experiment, all the spectral features are masked. This emphasizes importance of high quality samples to get by resonant Brillouin scattering a clear evidence of polariton formation.

\begin{figure}[tb]
\centering\includegraphics[width=0.8\columnwidth]{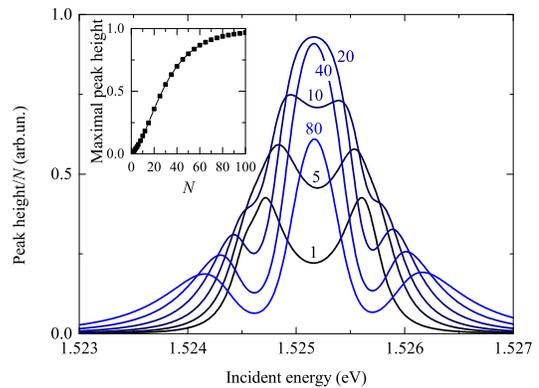}
\caption{Dependence of the scattering peak height 
on the incident photon energy for anti-Stokes scattering on the FLA2 mode.
Curves are plotted for different numbers of quantum wells in the structure,
the color of the line continuously evolves from black to blue with increasing value of $N$.
 Other parameters are indicated in the beginning of Sec.~\ref{sec:results}. 
For better presentation the spectra  were divided by the corresponding number of wells $N$. The inset shows the dependence of the maximum value of scattering peak height on $N$.}
\label{fig:N}
\end{figure}

We show in Fig.~\ref{fig:N} how the dependence on incident frequency of the FLA2 anti-Stokes scattering peak height calculated with $\hbar\Gamma=0.17~$meV varies with increase of the structure length.
While the main part of the Fig.~\ref{fig:N} presents the normalized peak height spectra, the inset shows the dependence of the maximal peak height on $N$, i.e., the maximum value of scattering intensity. The intensity monotonously increases with $N$. For very small $N$, the growth is quadratic in $N$. In this case the sample thickness is smaller than the light wavelength. Hence, the waves scattered from different wells are in phase and $R\propto |\Delta(0)|^{2}\propto N^{2}$ , see Eqs.~\eqref{eq:Rfinal},\eqref{eq:B}.
At large $N$  the intensity saturates because the sample thickness becomes larger than the attenuation length of the polaritons, which is finite due to the exciton absorption and due to the  polariton stop-band formation.

 For $N=1$, the curve has two maxima, corresponding to coincidence of the energies of incident or scattered photon with the energy of exciton in the single well. With increase of the quantum well number $N$, some dips appear in outer sides of the two peaks ($N=10$) and become more and more pronounced. The internal sides of the two peaks merge into a single larger structure at the central energy ($N=20$). At larger $N$, the spectrum assumes a three-peak form. 
A striking consequence of this evolution is that the scattering minima appear at the frequencies of the exciton resonances $\omega_0$ and $\omega_0-\bar\Omega$, reflecting  emergence of a polariton gap at the exciton resonance energy. Thus, the three-peak spectrum in this case is a specific consequence of the polariton formation while a two-peak spectrum corresponds to an exciton resonance in a single well. 
Coming back to Fig.~\ref{fig:Gamma}, corresponding to a number of QWs in the polariton regime, it becomes clear that increasing the damping washes out the features associated with the polariton formation: the gap becomes less pronounced while the variations of both the Brillouin shift and the linewidth broaden. For the smallest shown damping $\hbar\Gamma=0.05$~meV, an additional dip appears in the intensity curve for FLA2. In this case, the separation between the polariton gap structures related to the incoming and outgoing resonances becomes larger than $\Gamma$ and these features are spectrally separated. The distortion of the linewidth peaks for small $\Gamma$, Figs.~\ref{fig:Gamma}(c),and (d) is due to the fact that  the scattering peak shape deviates from Lorentzian and  peak width can not be straightforwardly defined.


Another interesting prediction of our model is the asymmetry between the peak width maxima corresponding to the resonances of incident and scattered light, cf. left and right peaks in Figs.~\ref{fig:Gamma}(c) and~(d).
While the peak corresponding to the incident wave resonance is the same for all situations, the one corresponding to the scattered wave resonance is either larger or smaller depending on the phonon branch [cf. Figs.~\ref{fig:Gamma}(c) and~(d), corresponding to FLA1 and FLA2 cases], and on whether it is Stokes or anti-Stokes process [cf. magenta lines/red circles and blue lines/black triangles  in Fig.~\ref{fig:fit}(e)]. This asymmetry becomes very large for small values of $\Gamma$ and remains observable in some of the experimental data, see Fig.~\ref{fig:fit}(e).
The effect cannot be explained by the simplified model of Ref.~\onlinecite{Jusserand2012} and looks surprising at a first sight owing to the identical contributions of the incident and scattered waves into the conservation laws Eq.~\eqref{eq:cases}.
The asymmetry originates from the fact that the scattering process depends on the joint density of states of the relevant phonon branch and the scattered polariton [that changes  when the slope of the phonon branch is inverted either going from FLA1 to FLA2, see Fig.~\ref{fig:map}(a), or from anti-Stokes to Stokes component], but does not depend on the density of states of incident wave.
Particularly, in long enough structures the scattering spectrum is governed by the term $\Delta(Q\init+Q\scat+k)$ in Eq.~\eqref{eq:B}. Analysis of this term yields the following approximate analytical expression for the scattering peak width as a function of the incident photon energy,
\begin{equation}
W(\omega\init)=\frac{\sqrt{\left(\gamma_{1}/dN\right)^{2}+[\gamma_{2}\Im (Q\init+Q\scat)]^{2}}}{1/s\mp \eta \Re (\rmd Q\scat/\rmd\omega\scat)}
\:.\label{eq:HWHM}
\end{equation}
Here, $\omega\scat$ should be found from the conservation laws and can be approximated as $\omega\scat=\omega\init\pm \bar \Omega$, the upper (lower) sign corresponds to the anti-Stokes (Stokes) scattering, parameter $\eta$ determines the slope of phonon branch and equals to $+1$ for LA and FLA2 scattering and to $-1$ for FLA1 scattering. The dimensionless parameters  $\gamma_{1}$ and $\gamma_{2}$ are determined approximating the half width at half maximum of  $|\Delta(K)|^{2}$ as a function of $\Re K$ by $\sqrt{(\gamma_1/dN)^2+(\gamma_2\Im K)^2}$. Best agreement with the exact calculation is obtained for $\gamma_{1}\approx 6$ and $\gamma_{2}\approx 2$.

The numerator of 
Eq.~\eqref{eq:HWHM} shows two contributions to the scattering peak width, describing the relaxation of the momentum conservation law due to the finite size of the structure ($\propto \gamma_{1}$ ) and due to the exciton absorption ($\propto \gamma_{2}$). The asymmetry between the incoming and outgoing resonances in the width spectrum stems from the denominator of Eq.~\eqref{eq:HWHM}, that is sensitive to the relative slope of the phonon and polariton dispersion curves. 

\section{Summary}\label{sec:summary}

\label{sec:summary} To summarize, we have developed a theory of the
Brillouin scattering of excitonic polaritons in the multiple-quantum-well
structure. Analytical results for the scattered light spectrum as function
of the incident light frequency and quantum well parameters have been
obtained.

The theory has been applied to describe the scattering experiment for the
superlattice, where the period is short as compared to the light wavelength.
The results are in perfect quantitative agreement with experimental data. 
We demonstrate, that the Brillouin shift spectrum follows from the exciton-polariton dispersion law in the infinite
superlattice, and weakly depends on the structure length. However, the width of the scattering peak is strongly sensitive to the number of quantum wells. It is
contributed by the exciton nonradiative damping and the uncertainty in the
scattered polariton wave vector in the structure of finite length. The calculations reveal pronounced asymmetry between the two maxima in the scattering peak width spectrum, corresponding to the exciton resonances for incident and scattered waves. The  asymmetry,  stemming from the finite ratio of
the sound speed and the polariton group velocity, differs for Stokes and anti-Stokes scattering and depends on the used folded phonon branch. The scattered intensity spectrum also qualitatively varies with structure length. While for a small number of quantum wells it has two peaks, corresponding to the incident and scattered resonances in single quantum wells, in long structures the spectrum has a three-peak form, that reflects formation of collective exciton-polaritonic modes.

The developed general theory of Brillouin scattering in multiple-quantum-well structures can be easily applied to Bragg periodic and aperiodic superlattices with interwell distances comparable with the light wavelength, provided that such demanding structure becomes sufficiently mature in terms of molecular beam epitaxy and available for Brillouin scattering experiments.

\acknowledgements
The authors acknowledge fruitful discussions with E.L.~Ivchenko, S.A.~Tarasenko, and A.~Fainstein. This work was supported by the RFBR, RF President Grant Nos.
MD-2062.2012.2 and NSh-5442.2012.2, EU projects POLAPHEN and SPANGL4Q, 
 the Region Ile-de-France in the framework of C'Nano IdF, the nanoscience competence center of Paris Region, and
the ``Dynasty'' Foundation.

\begin{thebibliography}{37}%
\makeatletter
\providecommand \@ifxundefined [1]{%
 \@ifx{#1\undefined}
}%
\providecommand \@ifnum [1]{%
 \ifnum #1\expandafter \@firstoftwo
 \else \expandafter \@secondoftwo
 \fi
}%
\providecommand \@ifx [1]{%
 \ifx #1\expandafter \@firstoftwo
 \else \expandafter \@secondoftwo
 \fi
}%
\providecommand \natexlab [1]{#1}%
\providecommand \enquote  [1]{``#1''}%
\providecommand \bibnamefont  [1]{#1}%
\providecommand \bibfnamefont [1]{#1}%
\providecommand \citenamefont [1]{#1}%
\providecommand \href@noop [0]{\@secondoftwo}%
\providecommand \href [0]{\begingroup \@sanitize@url \@href}%
\providecommand \@href[1]{\@@startlink{#1}\@@href}%
\providecommand \@@href[1]{\endgroup#1\@@endlink}%
\providecommand \@sanitize@url [0]{\catcode `\\12\catcode `\$12\catcode
  `\&12\catcode `\#12\catcode `\^12\catcode `\_12\catcode `\%12\relax}%
\providecommand \@@startlink[1]{}%
\providecommand \@@endlink[0]{}%
\providecommand \url  [0]{\begingroup\@sanitize@url \@url }%
\providecommand \@url [1]{\endgroup\@href {#1}{\urlprefix }}%
\providecommand \urlprefix  [0]{URL }%
\providecommand \Eprint [0]{\href }%
\providecommand \doibase [0]{http://dx.doi.org/}%
\providecommand \selectlanguage [0]{\@gobble}%
\providecommand \bibinfo  [0]{\@secondoftwo}%
\providecommand \bibfield  [0]{\@secondoftwo}%
\providecommand \translation [1]{[#1]}%
\providecommand \BibitemOpen [0]{}%
\providecommand \bibitemStop [0]{}%
\providecommand \bibitemNoStop [0]{.\EOS\space}%
\providecommand \EOS [0]{\spacefactor3000\relax}%
\providecommand \BibitemShut  [1]{\csname bibitem#1\endcsname}%
\let\auto@bib@innerbib\@empty
\bibitem [{\citenamefont {Hopfield}(1958)}]{hopfield58}%
  \BibitemOpen
  \bibfield  {author} {\bibinfo {author} {\bibfnamefont {J.~J.}\ \bibnamefont
  {Hopfield}},\ }\bibfield  {title} {\enquote {\bibinfo {title} {{T}heory of
  the {C}ontribution of {E}xcitons to the {C}omplex {D}ielectric {C}onstant of
  {C}rystals},}\ }\href {\doibase 10.1103/PhysRev.112.1555} {\bibfield
  {journal} {\bibinfo  {journal} {Phys. Rev.}\ }\textbf {\bibinfo {volume}
  {112}},\ \bibinfo {pages} {1555--1567} (\bibinfo {year} {1958})}\BibitemShut
  {NoStop}%
\bibitem [{\citenamefont {Kavokin}\ \emph {et~al.}(2006)\citenamefont
  {Kavokin}, \citenamefont {Baumberg}, \citenamefont {Malpuech},\ and\
  \citenamefont {Laussy}}]{kavbamalas}%
  \BibitemOpen
  \bibfield  {author} {\bibinfo {author} {\bibfnamefont {A.}~\bibnamefont
  {Kavokin}}, \bibinfo {author} {\bibfnamefont {J.J.}\ \bibnamefont
  {Baumberg}}, \bibinfo {author} {\bibfnamefont {G.}~\bibnamefont {Malpuech}},
  \ and\ \bibinfo {author} {\bibfnamefont {F.P.}\ \bibnamefont {Laussy}},\
  }\href@noop {} {\emph {\bibinfo {title} {{M}icrocavities}}}\ (\bibinfo
  {publisher} {Clarendon Press},\ \bibinfo {address} {Oxford},\ \bibinfo {year}
  {2006})\BibitemShut {NoStop}%
\bibitem [{\citenamefont {{Ivchenko}}\ \emph {et~al.}(1994)\citenamefont
  {{Ivchenko}}, \citenamefont {{Nesvizhskii}},\ and\ \citenamefont
  {{Jorda}}}]{Ivchenko1994}%
  \BibitemOpen
  \bibfield  {author} {\bibinfo {author} {\bibfnamefont {E.~L.}\ \bibnamefont
  {{Ivchenko}}}, \bibinfo {author} {\bibfnamefont {A.~I.}\ \bibnamefont
  {{Nesvizhskii}}}, \ and\ \bibinfo {author} {\bibfnamefont {S.}~\bibnamefont
  {{Jorda}}},\ }\bibfield  {title} {\enquote {\bibinfo {title} {{B}ragg
  reflection of light from quantum-well structures},}\ }\href@noop {}
  {\bibfield  {journal} {\bibinfo  {journal} {Phys. Solid State}\ }\textbf
  {\bibinfo {volume} {36}},\ \bibinfo {pages} {1156--1161} (\bibinfo {year}
  {1994})}\BibitemShut {NoStop}%
\bibitem [{\citenamefont {Ivchenko}(2005)}]{Ivchenko2005}%
  \BibitemOpen
  \bibfield  {author} {\bibinfo {author} {\bibfnamefont {E.~L.}\ \bibnamefont
  {Ivchenko}},\ }\href@noop {} {\emph {\bibinfo {title} {{O}ptical
  {S}pectroscopy of {S}emiconductor {N}anostructures}}}\ (\bibinfo  {publisher}
  {Alpha Science International},\ \bibinfo {address} {Harrow, UK},\ \bibinfo
  {year} {2005})\BibitemShut {NoStop}%
\bibitem [{\citenamefont {Ivchenko}\ \emph {et~al.}(2004)\citenamefont
  {Ivchenko}, \citenamefont {Voronov}, \citenamefont {Erementchouk},
  \citenamefont {Deych},\ and\ \citenamefont {Lisyansky}}]{Voronov2004}%
  \BibitemOpen
  \bibfield  {author} {\bibinfo {author} {\bibfnamefont {E.~L.}\ \bibnamefont
  {Ivchenko}}, \bibinfo {author} {\bibfnamefont {M.~M.}\ \bibnamefont
  {Voronov}}, \bibinfo {author} {\bibfnamefont {M.~V.}\ \bibnamefont
  {Erementchouk}}, \bibinfo {author} {\bibfnamefont {L.~I.}\ \bibnamefont
  {Deych}}, \ and\ \bibinfo {author} {\bibfnamefont {A.~A.}\ \bibnamefont
  {Lisyansky}},\ }\bibfield  {title} {\enquote {\bibinfo {title}
  {{M}ultiple-quantum-well-based photonic crystals with simple and compound
  elementary supercells},}\ }\href {\doibase 10.1103/PhysRevB.70.195106}
  {\bibfield  {journal} {\bibinfo  {journal} {Phys. Rev. B}\ }\textbf {\bibinfo
  {volume} {70}},\ \bibinfo {pages} {195106} (\bibinfo {year}
  {2004})}\BibitemShut {NoStop}%
\bibitem [{\citenamefont {Poddubny}\ \emph {et~al.}(2008)\citenamefont
  {Poddubny}, \citenamefont {Pilozzi}, \citenamefont {Voronov},\ and\
  \citenamefont {Ivchenko}}]{Poddubny2008prb}%
  \BibitemOpen
  \bibfield  {author} {\bibinfo {author} {\bibfnamefont {A.~N.}\ \bibnamefont
  {Poddubny}}, \bibinfo {author} {\bibfnamefont {L.}~\bibnamefont {Pilozzi}},
  \bibinfo {author} {\bibfnamefont {M.~M.}\ \bibnamefont {Voronov}}, \ and\
  \bibinfo {author} {\bibfnamefont {E.~L.}\ \bibnamefont {Ivchenko}},\
  }\bibfield  {title} {\enquote {\bibinfo {title} {{R}esonant
  $\mbox{F}$ibonacci quantum well structures in one dimension},}\ }\href
  {\doibase 10.1103/PhysRevB.77.113306} {\bibfield  {journal} {\bibinfo
  {journal} {Phys. Rev. B}\ }\textbf {\bibinfo {volume} {77}},\ \bibinfo
  {pages} {113306} (\bibinfo {year} {2008})}\BibitemShut {NoStop}%
\bibitem [{\citenamefont {{Poddubny}}\ and\ \citenamefont
  {{Ivchenko}}(2010)}]{Podd_Ivch}%
  \BibitemOpen
  \bibfield  {author} {\bibinfo {author} {\bibfnamefont {A.~N.}\ \bibnamefont
  {{Poddubny}}}\ and\ \bibinfo {author} {\bibfnamefont {E.~L.}\ \bibnamefont
  {{Ivchenko}}},\ }\bibfield  {title} {\enquote {\bibinfo {title} {{P}hotonic
  quasicrystalline and aperiodic structures},}\ }\href {\doibase
  10.1016/j.physe.2010.02.020} {\bibfield  {journal} {\bibinfo  {journal}
  {Physica E}\ }\textbf {\bibinfo {volume} {42}},\ \bibinfo {pages}
  {1871--1895} (\bibinfo {year} {2010})}\BibitemShut {NoStop}%
\bibitem [{\citenamefont {Averkiev}\ \emph {et~al.}(2012)\citenamefont
  {Averkiev}, \citenamefont {Glazov},\ and\ \citenamefont
  {Voronov}}]{Averkiev2012}%
  \BibitemOpen
  \bibfield  {author} {\bibinfo {author} {\bibfnamefont {N.S.}\ \bibnamefont
  {Averkiev}}, \bibinfo {author} {\bibfnamefont {M.M.}\ \bibnamefont {Glazov}},
  \ and\ \bibinfo {author} {\bibfnamefont {M.M.}\ \bibnamefont {Voronov}},\
  }\bibfield  {title} {\enquote {\bibinfo {title} {{F}ermi-edge polaritons in
  {B}ragg multiple-quantum-well structures},}\ }\href {\doibase
  10.1016/j.ssc.2011.12.002} {\bibfield  {journal} {\bibinfo  {journal} {Solid
  State Comm.}\ }\textbf {\bibinfo {volume} {152}},\ \bibinfo {pages} {395 --
  398} (\bibinfo {year} {2012})}\BibitemShut {NoStop}%
\bibitem [{\citenamefont {Poshakinskiy}\ \emph {et~al.}(2012)\citenamefont
  {Poshakinskiy}, \citenamefont {Poddubny},\ and\ \citenamefont
  {Tarasenko}}]{Poshakinskiy2012}%
  \BibitemOpen
  \bibfield  {author} {\bibinfo {author} {\bibfnamefont {A.~V.}\ \bibnamefont
  {Poshakinskiy}}, \bibinfo {author} {\bibfnamefont {A.~N.}\ \bibnamefont
  {Poddubny}}, \ and\ \bibinfo {author} {\bibfnamefont {S.~A.}\ \bibnamefont
  {Tarasenko}},\ }\bibfield  {title} {\enquote {\bibinfo {title} {Reflection of
  short polarized optical pulses from periodic and aperiodic multiple quantum
  well structures},}\ }\href {\doibase 10.1103/PhysRevB.86.205304} {\bibfield
  {journal} {\bibinfo  {journal} {Phys. Rev. B}\ }\textbf {\bibinfo {volume}
  {86}},\ \bibinfo {pages} {205304} (\bibinfo {year} {2012})}\BibitemShut
  {NoStop}%
\bibitem [{\citenamefont {Prineas}\ \emph
  {et~al.}(2006{\natexlab{a}})\citenamefont {Prineas}, \citenamefont {Cao},
  \citenamefont {Yildirim}, \citenamefont {Johnston},\ and\ \citenamefont
  {Reddy}}]{prineas2006}%
  \BibitemOpen
  \bibfield  {author} {\bibinfo {author} {\bibfnamefont {J.~P.}\ \bibnamefont
  {Prineas}}, \bibinfo {author} {\bibfnamefont {C.}~\bibnamefont {Cao}},
  \bibinfo {author} {\bibfnamefont {M.}~\bibnamefont {Yildirim}}, \bibinfo
  {author} {\bibfnamefont {W.}~\bibnamefont {Johnston}}, \ and\ \bibinfo
  {author} {\bibfnamefont {M.}~\bibnamefont {Reddy}},\ }\bibfield  {title}
  {\enquote {\bibinfo {title} {{R}esonant photonic band gap structures realized
  from molecular-beam-epitaxially grown {I}n{G}a{A}s/{G}a{A}s {B}ragg-spaced
  quantum wells},}\ }\href {\doibase 10.1063/1.2234814} {\bibfield  {journal}
  {\bibinfo  {journal} {J. Appl. Phys.}\ }\textbf {\bibinfo {volume} {100}},\
  \bibinfo {eid} {063101} (\bibinfo {year} {2006}{\natexlab{a}})}\BibitemShut
  {NoStop}%
\bibitem [{\citenamefont {Prineas}\ \emph
  {et~al.}(2006{\natexlab{b}})\citenamefont {Prineas}, \citenamefont
  {Johnston}, \citenamefont {Yildirim}, \citenamefont {Zhao},\ and\
  \citenamefont {Smirl}}]{prineas2006apl}%
  \BibitemOpen
  \bibfield  {author} {\bibinfo {author} {\bibfnamefont {J.~P.}\ \bibnamefont
  {Prineas}}, \bibinfo {author} {\bibfnamefont {W.~J.}\ \bibnamefont
  {Johnston}}, \bibinfo {author} {\bibfnamefont {M.}~\bibnamefont {Yildirim}},
  \bibinfo {author} {\bibfnamefont {J.}~\bibnamefont {Zhao}}, \ and\ \bibinfo
  {author} {\bibfnamefont {Arthur~L.}\ \bibnamefont {Smirl}},\ }\bibfield
  {title} {\enquote {\bibinfo {title} {{T}unable slow light in {B}ragg-spaced
  quantum wells},}\ }\href {\doibase 10.1063/1.2403927} {\bibfield  {journal}
  {\bibinfo  {journal} {Appl. Phys. Lett.}\ }\textbf {\bibinfo {volume} {89}},\
  \bibinfo {eid} {241106} (\bibinfo {year} {2006}{\natexlab{b}})}\BibitemShut
  {NoStop}%
\bibitem [{\citenamefont {Hendrickson}\ \emph {et~al.}(2008)\citenamefont
  {Hendrickson}, \citenamefont {Richards}, \citenamefont {Sweet}, \citenamefont
  {Khitrova}, \citenamefont {Poddubny}, \citenamefont {Ivchenko}, \citenamefont
  {Wegener},\ and\ \citenamefont {Gibbs}}]{Hendrickson2008}%
  \BibitemOpen
  \bibfield  {author} {\bibinfo {author} {\bibfnamefont {J.}~\bibnamefont
  {Hendrickson}}, \bibinfo {author} {\bibfnamefont {B.~C.}\ \bibnamefont
  {Richards}}, \bibinfo {author} {\bibfnamefont {J.}~\bibnamefont {Sweet}},
  \bibinfo {author} {\bibfnamefont {G.}~\bibnamefont {Khitrova}}, \bibinfo
  {author} {\bibfnamefont {A.~N.}\ \bibnamefont {Poddubny}}, \bibinfo {author}
  {\bibfnamefont {E.~L.}\ \bibnamefont {Ivchenko}}, \bibinfo {author}
  {\bibfnamefont {M.}~\bibnamefont {Wegener}}, \ and\ \bibinfo {author}
  {\bibfnamefont {H.~M.}\ \bibnamefont {Gibbs}},\ }\bibfield  {title} {\enquote
  {\bibinfo {title} {{E}xcitonic polaritons in $\mbox{F}$ibonacci
  quasicrystals},}\ }\href {\doibase 10.1364/OE.16.015382} {\bibfield
  {journal} {\bibinfo  {journal} {Opt. Express}\ }\textbf {\bibinfo {volume}
  {16}},\ \bibinfo {pages} {15382--15387} (\bibinfo {year} {2008})}\BibitemShut
  {NoStop}%
\bibitem [{\citenamefont {{Goldberg}}\ \emph {et~al.}(2009)\citenamefont
  {{Goldberg}}, \citenamefont {{Deych}}, \citenamefont {{Lisyansky}},
  \citenamefont {{Shi}}, \citenamefont {{Menon}}, \citenamefont {{Tokranov}},
  \citenamefont {{Yakimov}},\ and\ \citenamefont
  {{Oktyabrsky}}}]{Goldberg2009}%
  \BibitemOpen
  \bibfield  {author} {\bibinfo {author} {\bibfnamefont {D.}~\bibnamefont
  {{Goldberg}}}, \bibinfo {author} {\bibfnamefont {L.~I.}\ \bibnamefont
  {{Deych}}}, \bibinfo {author} {\bibfnamefont {A.~A.}\ \bibnamefont
  {{Lisyansky}}}, \bibinfo {author} {\bibfnamefont {Z.}~\bibnamefont {{Shi}}},
  \bibinfo {author} {\bibfnamefont {V.~M.}\ \bibnamefont {{Menon}}}, \bibinfo
  {author} {\bibfnamefont {V.}~\bibnamefont {{Tokranov}}}, \bibinfo {author}
  {\bibfnamefont {M.}~\bibnamefont {{Yakimov}}}, \ and\ \bibinfo {author}
  {\bibfnamefont {S.}~\bibnamefont {{Oktyabrsky}}},\ }\bibfield  {title}
  {\enquote {\bibinfo {title} {{E}xciton-lattice polaritons in
  multiple-quantum-well-based photonic crystals},}\ }\href {\doibase
  10.1038/nphoton.2009.190} {\bibfield  {journal} {\bibinfo  {journal} {Nature
  Photonics}\ }\textbf {\bibinfo {volume} {3}},\ \bibinfo {pages} {662--666}
  (\bibinfo {year} {2009})}\BibitemShut {NoStop}%
\bibitem [{\citenamefont {Askitopoulos}\ \emph {et~al.}(2011)\citenamefont
  {Askitopoulos}, \citenamefont {Mouchliadis}, \citenamefont {Iorsh},
  \citenamefont {Christmann}, \citenamefont {Baumberg}, \citenamefont
  {Kaliteevski}, \citenamefont {Hatzopoulos},\ and\ \citenamefont
  {Savvidis}}]{iorsh2011}%
  \BibitemOpen
  \bibfield  {author} {\bibinfo {author} {\bibfnamefont {A.}~\bibnamefont
  {Askitopoulos}}, \bibinfo {author} {\bibfnamefont {L.}~\bibnamefont
  {Mouchliadis}}, \bibinfo {author} {\bibfnamefont {I.}~\bibnamefont {Iorsh}},
  \bibinfo {author} {\bibfnamefont {G.}~\bibnamefont {Christmann}}, \bibinfo
  {author} {\bibfnamefont {J.~J.}\ \bibnamefont {Baumberg}}, \bibinfo {author}
  {\bibfnamefont {M.~A.}\ \bibnamefont {Kaliteevski}}, \bibinfo {author}
  {\bibfnamefont {Z.}~\bibnamefont {Hatzopoulos}}, \ and\ \bibinfo {author}
  {\bibfnamefont {P.~G.}\ \bibnamefont {Savvidis}},\ }\bibfield  {title}
  {\enquote {\bibinfo {title} {{B}ragg {P}olaritons: {S}trong {C}oupling and
  {A}mplification in an {U}nfolded {M}icrocavity},}\ }\href {\doibase
  10.1103/PhysRevLett.106.076401} {\bibfield  {journal} {\bibinfo  {journal}
  {Phys. Rev. Lett.}\ }\textbf {\bibinfo {volume} {106}},\ \bibinfo {pages}
  {076401} (\bibinfo {year} {2011})}\BibitemShut {NoStop}%
\bibitem [{\citenamefont {Chaldyshev}\ \emph
  {et~al.}(2011{\natexlab{a}})\citenamefont {Chaldyshev}, \citenamefont {Chen},
  \citenamefont {Poddubny}, \citenamefont {Vasil'ev},\ and\ \citenamefont
  {Liu}}]{chaldyshev2011}%
  \BibitemOpen
  \bibfield  {author} {\bibinfo {author} {\bibfnamefont {V.~V.}\ \bibnamefont
  {Chaldyshev}}, \bibinfo {author} {\bibfnamefont {Yuechao}\ \bibnamefont
  {Chen}}, \bibinfo {author} {\bibfnamefont {A.~N.}\ \bibnamefont {Poddubny}},
  \bibinfo {author} {\bibfnamefont {A.~P.}\ \bibnamefont {Vasil'ev}}, \ and\
  \bibinfo {author} {\bibfnamefont {Zhiheng}\ \bibnamefont {Liu}},\ }\bibfield
  {title} {\enquote {\bibinfo {title} {{R}esonant optical reflection by a
  periodic system of the quantum well excitons at the second quantum state},}\
  }\href {\doibase 10.1063/1.3554429} {\bibfield  {journal} {\bibinfo
  {journal} {Appl. Phys. Lett.}\ }\textbf {\bibinfo {volume} {98}},\ \bibinfo
  {eid} {073112} (\bibinfo {year} {2011}{\natexlab{a}})}\BibitemShut {NoStop}%
\bibitem [{\citenamefont {Chaldyshev}\ \emph
  {et~al.}(2011{\natexlab{b}})\citenamefont {Chaldyshev}, \citenamefont
  {Bolshakov}, \citenamefont {Zavarin}, \citenamefont {Sakharov}, \citenamefont
  {Lundin}, \citenamefont {Tsatsulnikov}, \citenamefont {Yagovkina},
  \citenamefont {Kim},\ and\ \citenamefont {Park}}]{chaldyshev2011b}%
  \BibitemOpen
  \bibfield  {author} {\bibinfo {author} {\bibfnamefont {V.~V.}\ \bibnamefont
  {Chaldyshev}}, \bibinfo {author} {\bibfnamefont {A.~S.}\ \bibnamefont
  {Bolshakov}}, \bibinfo {author} {\bibfnamefont {E.~E.}\ \bibnamefont
  {Zavarin}}, \bibinfo {author} {\bibfnamefont {A.~V.}\ \bibnamefont
  {Sakharov}}, \bibinfo {author} {\bibfnamefont {W.~V.}\ \bibnamefont
  {Lundin}}, \bibinfo {author} {\bibfnamefont {A.~F.}\ \bibnamefont
  {Tsatsulnikov}}, \bibinfo {author} {\bibfnamefont {M.~A.}\ \bibnamefont
  {Yagovkina}}, \bibinfo {author} {\bibfnamefont {Taek}\ \bibnamefont {Kim}}, \
  and\ \bibinfo {author} {\bibfnamefont {Youngsoo}\ \bibnamefont {Park}},\
  }\bibfield  {title} {\enquote {\bibinfo {title} {{O}ptical lattices of
  {I}n{G}a{N} quantum well excitons},}\ }\href {\doibase 10.1063/1.3670499}
  {\bibfield  {journal} {\bibinfo  {journal} {Appl. Phys. Lett.}\ }\textbf
  {\bibinfo {volume} {99}},\ \bibinfo {eid} {251103} (\bibinfo {year}
  {2011}{\natexlab{b}})}\BibitemShut {NoStop}%
\bibitem [{\citenamefont {Chaldyshev}\ \emph {et~al.}(2012)\citenamefont
  {Chaldyshev}, \citenamefont {Kundelev}, \citenamefont {Nikitina},
  \citenamefont {Egorov},\ and\ \citenamefont {Gorbatsevich}}]{Chaldyshev2012}%
  \BibitemOpen
  \bibfield  {author} {\bibinfo {author} {\bibfnamefont {V.~V.}\ \bibnamefont
  {Chaldyshev}}, \bibinfo {author} {\bibfnamefont {E.~V.}\ \bibnamefont
  {Kundelev}}, \bibinfo {author} {\bibfnamefont {E.~V.}\ \bibnamefont
  {Nikitina}}, \bibinfo {author} {\bibfnamefont {A.~Yu.}\ \bibnamefont
  {Egorov}}, \ and\ \bibinfo {author} {\bibfnamefont {A.~A.}\ \bibnamefont
  {Gorbatsevich}},\ }\bibfield  {title} {\enquote {\bibinfo {title} {Resonance
  reflection of light by a periodic system of excitons in {G}a{A}s/{A}l{G}a{A}s
  quantum wells},}\ }\href {\doibase 10.1134/S1063782612080052} {\bibfield
  {journal} {\bibinfo  {journal} {Semiconductors}\ }\textbf {\bibinfo {volume}
  {46}},\ \bibinfo {pages} {1016--1019} (\bibinfo {year} {2012})}\BibitemShut
  {NoStop}%
\bibitem [{\citenamefont {Poddubny}\ and\ \citenamefont
  {Ivchenko}(2013)}]{Ivchenko2013}%
  \BibitemOpen
  \bibfield  {author} {\bibinfo {author} {\bibfnamefont {A.N.}\ \bibnamefont
  {Poddubny}}\ and\ \bibinfo {author} {\bibfnamefont {E.L.}\ \bibnamefont
  {Ivchenko}},\ }\bibfield  {title} {\enquote {\bibinfo {title} {Resonant
  diffraction of electromagnetic waves from solids (a review)},}\ }\href
  {\doibase 10.1134/S1063783413050120} {\bibfield  {journal} {\bibinfo
  {journal} {Phys. Solid State}\ }\textbf {\bibinfo {volume} {55}},\ \bibinfo
  {pages} {905--923} (\bibinfo {year} {2013})}\BibitemShut {NoStop}%
\bibitem [{\citenamefont {Weisbuch}\ and\ \citenamefont
  {Ulbrich}(1982)}]{Weishuch1982}%
  \BibitemOpen
  \bibfield  {author} {\bibinfo {author} {\bibfnamefont {C.}~\bibnamefont
  {Weisbuch}}\ and\ \bibinfo {author} {\bibfnamefont {R.G.}\ \bibnamefont
  {Ulbrich}},\ }\href@noop {} {\emph {\bibinfo {title} {{L}ight {S}cattering in
  {S}olids {III}}}},\ edited by\ \bibinfo {editor} {\bibfnamefont
  {{M.}}~\bibnamefont {{C}ardona}}\ and\ \bibinfo {editor} {\bibfnamefont
  {{G}.}~\bibnamefont {{G}\"untherodt}}\ (\bibinfo  {publisher} {Springer},\
  \bibinfo {address} {Berlin},\ \bibinfo {year} {1982})\ p.\ \bibinfo {pages}
  {207}\BibitemShut {NoStop}%
\bibitem [{\citenamefont {Yu}\ and\ \citenamefont
  {Evangelisti}(1979)}]{yu1979}%
  \BibitemOpen
  \bibfield  {author} {\bibinfo {author} {\bibfnamefont {Peter~Y.}\
  \bibnamefont {Yu}}\ and\ \bibinfo {author} {\bibfnamefont {F.}~\bibnamefont
  {Evangelisti}},\ }\bibfield  {title} {\enquote {\bibinfo {title} {{B}rillouin
  {S}cattering {E}fficiencies of {E}xciton {P}olaritons and the {A}dditional
  {B}oundary {C}onditions in {C}d{S}},}\ }\href {\doibase
  10.1103/PhysRevLett.42.1642} {\bibfield  {journal} {\bibinfo  {journal}
  {Phys. Rev. Lett.}\ }\textbf {\bibinfo {volume} {42}},\ \bibinfo {pages}
  {1642--1645} (\bibinfo {year} {1979})}\BibitemShut {NoStop}%
\bibitem [{\citenamefont {Bendow}\ and\ \citenamefont
  {Birman}(1970)}]{Bendow1970}%
  \BibitemOpen
  \bibfield  {author} {\bibinfo {author} {\bibfnamefont {Bernard}\ \bibnamefont
  {Bendow}}\ and\ \bibinfo {author} {\bibfnamefont {Joseph~L.}\ \bibnamefont
  {Birman}},\ }\bibfield  {title} {\enquote {\bibinfo {title} {{P}olariton
  {T}heory of {R}esonance {R}aman {S}cattering in {I}nsulating {C}rystals},}\
  }\href {\doibase 10.1103/PhysRevB.1.1678} {\bibfield  {journal} {\bibinfo
  {journal} {Phys. Rev. B}\ }\textbf {\bibinfo {volume} {1}},\ \bibinfo {pages}
  {1678--1686} (\bibinfo {year} {1970})}\BibitemShut {NoStop}%
\bibitem [{\citenamefont {Bendow}(1970)}]{Bendow1970b}%
  \BibitemOpen
  \bibfield  {author} {\bibinfo {author} {\bibfnamefont {Bernard}\ \bibnamefont
  {Bendow}},\ }\bibfield  {title} {\enquote {\bibinfo {title} {{P}olariton
  {T}heory of {R}aman {S}cattering in {I}nsulating {C}rystals. {II}.}}\ }\href
  {\doibase 10.1103/PhysRevB.2.5051} {\bibfield  {journal} {\bibinfo  {journal}
  {Phys. Rev. B}\ }\textbf {\bibinfo {volume} {2}},\ \bibinfo {pages}
  {5051--5062} (\bibinfo {year} {1970})}\BibitemShut {NoStop}%
\bibitem [{\citenamefont {Zeyher}\ \emph {et~al.}(1974)\citenamefont {Zeyher},
  \citenamefont {Ting},\ and\ \citenamefont {Birman}}]{Zeyher1974}%
  \BibitemOpen
  \bibfield  {author} {\bibinfo {author} {\bibfnamefont {Roland}\ \bibnamefont
  {Zeyher}}, \bibinfo {author} {\bibfnamefont {Chin-Sen}\ \bibnamefont {Ting}},
  \ and\ \bibinfo {author} {\bibfnamefont {Joseph~L.}\ \bibnamefont {Birman}},\
  }\bibfield  {title} {\enquote {\bibinfo {title} {Polariton theory of
  first-order {R}aman scattering in finite crystals for transparent and
  absorbing frequency regions},}\ }\href {\doibase 10.1103/PhysRevB.10.1725}
  {\bibfield  {journal} {\bibinfo  {journal} {Phys. Rev. B}\ }\textbf {\bibinfo
  {volume} {10}},\ \bibinfo {pages} {1725--1740} (\bibinfo {year}
  {1974})}\BibitemShut {NoStop}%
\bibitem [{\citenamefont {Matsushita}\ \emph {et~al.}(1984)\citenamefont
  {Matsushita}, \citenamefont {Wicksted},\ and\ \citenamefont
  {Cummins}}]{Matsushita1984}%
  \BibitemOpen
  \bibfield  {author} {\bibinfo {author} {\bibfnamefont {M.}~\bibnamefont
  {Matsushita}}, \bibinfo {author} {\bibfnamefont {J.}~\bibnamefont
  {Wicksted}}, \ and\ \bibinfo {author} {\bibfnamefont {H.~Z.}\ \bibnamefont
  {Cummins}},\ }\bibfield  {title} {\enquote {\bibinfo {title} {Resonant
  brillouin scattering in cds. ii. theory},}\ }\href {\doibase
  10.1103/PhysRevB.29.3362} {\bibfield  {journal} {\bibinfo  {journal} {Phys.
  Rev. B}\ }\textbf {\bibinfo {volume} {29}},\ \bibinfo {pages} {3362--3381}
  (\bibinfo {year} {1984})}\BibitemShut {NoStop}%
\bibitem [{\citenamefont {Matsushita}\ and\ \citenamefont
  {Nakayama}(1984)}]{Matsushita1974b}%
  \BibitemOpen
  \bibfield  {author} {\bibinfo {author} {\bibfnamefont {M.}~\bibnamefont
  {Matsushita}}\ and\ \bibinfo {author} {\bibfnamefont {M.}~\bibnamefont
  {Nakayama}},\ }\bibfield  {title} {\enquote {\bibinfo {title} {Theory of
  resonant light scattering through exciton-polaritons},}\ }\href {\doibase
  10.1103/PhysRevB.30.2074} {\bibfield  {journal} {\bibinfo  {journal} {Phys.
  Rev. B}\ }\textbf {\bibinfo {volume} {30}},\ \bibinfo {pages} {2074--2083}
  (\bibinfo {year} {1984})}\BibitemShut {NoStop}%
\bibitem [{\citenamefont {Fainstein}\ \emph {et~al.}(1997)\citenamefont
  {Fainstein}, \citenamefont {Jusserand},\ and\ \citenamefont
  {Thierry-Mieg}}]{Fainstein1997}%
  \BibitemOpen
  \bibfield  {author} {\bibinfo {author} {\bibfnamefont {A.}~\bibnamefont
  {Fainstein}}, \bibinfo {author} {\bibfnamefont {B.}~\bibnamefont
  {Jusserand}}, \ and\ \bibinfo {author} {\bibfnamefont {V.}~\bibnamefont
  {Thierry-Mieg}},\ }\bibfield  {title} {\enquote {\bibinfo {title}
  {Cavity-polariton mediated resonant {R}aman scattering},}\ }\href {\doibase
  10.1103/PhysRevLett.78.1576} {\bibfield  {journal} {\bibinfo  {journal}
  {Phys. Rev. Lett.}\ }\textbf {\bibinfo {volume} {78}},\ \bibinfo {pages}
  {1576--1579} (\bibinfo {year} {1997})}\BibitemShut {NoStop}%
\bibitem [{\citenamefont {Bruchhausen}\ \emph {et~al.}(2008)\citenamefont
  {Bruchhausen}, \citenamefont {Hilario}, \citenamefont {Aligia}, \citenamefont
  {Lobos}, \citenamefont {Fainstein}, \citenamefont {Jusserand},\ and\
  \citenamefont {Andr\'e}}]{Bruchhausen2008}%
  \BibitemOpen
  \bibfield  {author} {\bibinfo {author} {\bibfnamefont {A.}~\bibnamefont
  {Bruchhausen}}, \bibinfo {author} {\bibfnamefont {L.~M.~Le\'on}\ \bibnamefont
  {Hilario}}, \bibinfo {author} {\bibfnamefont {A.~A.}\ \bibnamefont {Aligia}},
  \bibinfo {author} {\bibfnamefont {A.~M.}\ \bibnamefont {Lobos}}, \bibinfo
  {author} {\bibfnamefont {A.}~\bibnamefont {Fainstein}}, \bibinfo {author}
  {\bibfnamefont {B.}~\bibnamefont {Jusserand}}, \ and\ \bibinfo {author}
  {\bibfnamefont {R.}~\bibnamefont {Andr\'e}},\ }\bibfield  {title} {\enquote
  {\bibinfo {title} {Microcavity exciton-polariton mediated {R}aman scattering:
  Experiments and theory},}\ }\href {\doibase 10.1103/PhysRevB.78.125326}
  {\bibfield  {journal} {\bibinfo  {journal} {Phys. Rev. B}\ }\textbf {\bibinfo
  {volume} {78}},\ \bibinfo {pages} {125326} (\bibinfo {year}
  {2008})}\BibitemShut {NoStop}%
\bibitem [{\citenamefont {Rozas}\ \emph {et~al.}()\citenamefont {Rozas},
  \citenamefont {Bruchhausen}, \citenamefont {Fainstein}, \citenamefont
  {Jusserand},\ and\ \citenamefont {Lema{\^\i}tre}}]{Jusserand2011}%
  \BibitemOpen
  \bibfield  {author} {\bibinfo {author} {\bibfnamefont {G.}~\bibnamefont
  {Rozas}}, \bibinfo {author} {\bibfnamefont {A.}~\bibnamefont {Bruchhausen}},
  \bibinfo {author} {\bibfnamefont {A.}~\bibnamefont {Fainstein}}, \bibinfo
  {author} {\bibfnamefont {B.}~\bibnamefont {Jusserand}}, \ and\ \bibinfo
  {author} {\bibfnamefont {A.}~\bibnamefont {Lema{\^\i}tre}},\ }\href@noop {}
  {}\bibinfo {note} {{u}npublished}\BibitemShut {NoStop}%
\bibitem [{\citenamefont {Jusserand}\ and\ \citenamefont
  {Cardona}(1989)}]{Jusserand1989}%
  \BibitemOpen
  \bibfield  {author} {\bibinfo {author} {\bibfnamefont {B.}~\bibnamefont
  {Jusserand}}\ and\ \bibinfo {author} {\bibfnamefont {M.}~\bibnamefont
  {Cardona}},\ }\enquote {\bibinfo {title} {Light scattering in solids {V}},}\
  \ (\bibinfo  {publisher} {Springer},\ \bibinfo {year} {1989})\ p.~\bibinfo
  {pages} {49}\BibitemShut {NoStop}%
\bibitem [{\citenamefont {Jusserand}\ \emph {et~al.}(1987)\citenamefont
  {Jusserand}, \citenamefont {Paquet}, \citenamefont {Mollot}, \citenamefont
  {Alexandre},\ and\ \citenamefont {Le~Roux}}]{Jusserand1987}%
  \BibitemOpen
  \bibfield  {author} {\bibinfo {author} {\bibfnamefont {B.}~\bibnamefont
  {Jusserand}}, \bibinfo {author} {\bibfnamefont {D.}~\bibnamefont {Paquet}},
  \bibinfo {author} {\bibfnamefont {F.}~\bibnamefont {Mollot}}, \bibinfo
  {author} {\bibfnamefont {F.}~\bibnamefont {Alexandre}}, \ and\ \bibinfo
  {author} {\bibfnamefont {G.}~\bibnamefont {Le~Roux}},\ }\bibfield  {title}
  {\enquote {\bibinfo {title} {Influence of the supercell structure on the
  folded acoustical {R}aman line intensities in superlattices},}\ }\href
  {\doibase 10.1103/PhysRevB.35.2808} {\bibfield  {journal} {\bibinfo
  {journal} {Phys. Rev. B}\ }\textbf {\bibinfo {volume} {35}},\ \bibinfo
  {pages} {2808--2817} (\bibinfo {year} {1987})}\BibitemShut {NoStop}%
\bibitem [{\citenamefont {He}\ \emph {et~al.}(1988)\citenamefont {He},
  \citenamefont {Djafari-Rouhani},\ and\ \citenamefont {Sapriel}}]{He1988}%
  \BibitemOpen
  \bibfield  {author} {\bibinfo {author} {\bibfnamefont {Jianjun}\ \bibnamefont
  {He}}, \bibinfo {author} {\bibfnamefont {Bahram}\ \bibnamefont
  {Djafari-Rouhani}}, \ and\ \bibinfo {author} {\bibfnamefont {Jacques}\
  \bibnamefont {Sapriel}},\ }\bibfield  {title} {\enquote {\bibinfo {title}
  {{T}heory of light scattering by longitudinal-acoustic phonons in
  superlattices},}\ }\href {\doibase 10.1103/PhysRevB.37.4086} {\bibfield
  {journal} {\bibinfo  {journal} {Phys. Rev. B}\ }\textbf {\bibinfo {volume}
  {37}},\ \bibinfo {pages} {4086--4098} (\bibinfo {year} {1988})}\BibitemShut
  {NoStop}%
\bibitem [{\citenamefont {Jusserand}(2013)}]{Jusserand2013}%
  \BibitemOpen
  \bibfield  {author} {\bibinfo {author} {\bibfnamefont {B.}~\bibnamefont
  {Jusserand}},\ }\bibfield  {title} {\enquote {\bibinfo {title} {Selective
  resonant interaction between confined excitons and folded acoustic phonons in
  {GaAs/AlAs} multi-quantum wells},}\ }\href {\doibase
  http://dx.doi.org/10.1063/1.4817647} {\bibfield  {journal} {\bibinfo
  {journal} {Appl. Phys. Lett.}\ }\textbf {\bibinfo {volume} {103}},\ \bibinfo
  {eid} {093112} (\bibinfo {year} {2013})}\BibitemShut {NoStop}%
\bibitem [{\citenamefont {Jusserand}\ \emph {et~al.}(2012)\citenamefont
  {Jusserand}, \citenamefont {Fainstein}, \citenamefont {Ferreira},
  \citenamefont {Majrab},\ and\ \citenamefont {Lemaitre}}]{Jusserand2012}%
  \BibitemOpen
  \bibfield  {author} {\bibinfo {author} {\bibfnamefont {B.}~\bibnamefont
  {Jusserand}}, \bibinfo {author} {\bibfnamefont {A.}~\bibnamefont
  {Fainstein}}, \bibinfo {author} {\bibfnamefont {R.}~\bibnamefont {Ferreira}},
  \bibinfo {author} {\bibfnamefont {S.}~\bibnamefont {Majrab}}, \ and\ \bibinfo
  {author} {\bibfnamefont {A.}~\bibnamefont {Lemaitre}},\ }\bibfield  {title}
  {\enquote {\bibinfo {title} {{D}ispersion and damping of multiple
  quantum-well polaritons from resonant {B}rillouin scattering by folded
  acoustic modes},}\ }\href {\doibase 10.1103/PhysRevB.85.041302} {\bibfield
  {journal} {\bibinfo  {journal} {Phys. Rev. B}\ }\textbf {\bibinfo {volume}
  {85}},\ \bibinfo {pages} {041302} (\bibinfo {year} {2012})}\BibitemShut
  {NoStop}%
\bibitem [{\citenamefont {Yu}\ and\ \citenamefont {Cardona}(2010)}]{Cardona}%
  \BibitemOpen
  \bibfield  {author} {\bibinfo {author} {\bibfnamefont {P.Y.}\ \bibnamefont
  {Yu}}\ and\ \bibinfo {author} {\bibfnamefont {M.}~\bibnamefont {Cardona}},\
  }\href {http://books.google.se/books?id=5aBuKYBT\_hsC} {\emph {\bibinfo
  {title} {{F}undamentals of {S}emiconductors}}},\ Graduate texts in physics\ (\bibinfo  {publisher}
  {Springer},\ \bibinfo {year} {2010})\BibitemShut {NoStop}%
\bibitem [{\citenamefont {{Voronov}}\ \emph {et~al.}(2007)\citenamefont
  {{Voronov}}, \citenamefont {{Ivchenko}}, \citenamefont {{Erementchouk}},
  \citenamefont {{Deych}},\ and\ \citenamefont {{Lisyansky}}}]{Voronov2007}%
  \BibitemOpen
  \bibfield  {author} {\bibinfo {author} {\bibfnamefont {M.}~\bibnamefont
  {{Voronov}}}, \bibinfo {author} {\bibfnamefont {E.}~\bibnamefont
  {{Ivchenko}}}, \bibinfo {author} {\bibfnamefont {M.}~\bibnamefont
  {{Erementchouk}}}, \bibinfo {author} {\bibfnamefont {L.}~\bibnamefont
  {{Deych}}}, \ and\ \bibinfo {author} {\bibfnamefont {A.}~\bibnamefont
  {{Lisyansky}}},\ }\bibfield  {title} {\enquote {\bibinfo {title}
  {{P}hotoluminescence spectroscopy of one-dimensional resonant photonic
  crystals},}\ }\href {\doibase 10.1016/j.jlumin.2006.08.015} {\bibfield
  {journal} {\bibinfo  {journal} {J. of Luminescence}\ }\textbf {\bibinfo
  {volume} {125}},\ \bibinfo {pages} {112--117} (\bibinfo {year}
  {2007})}\BibitemShut {NoStop}%
\bibitem [{\citenamefont {Ivchenko}(1991)}]{Ivchenko1991}%
  \BibitemOpen
  \bibfield  {author} {\bibinfo {author} {\bibfnamefont {E.~L.}\ \bibnamefont
  {Ivchenko}},\ }\bibfield  {title} {\enquote {\bibinfo {title} {{E}xcitonic
  polaritons in periodic quantum-well structures},}\ }\href@noop {} {\bibfield
  {journal} {\bibinfo  {journal} {Sov. Phys. Sol. State}\ }\textbf {\bibinfo
  {volume} {33}},\ \bibinfo {pages} {1344--1346} (\bibinfo {year}
  {1991})}\BibitemShut {NoStop}%
\bibitem [{\citenamefont {Brenig}\ \emph {et~al.}(1972)\citenamefont {Brenig},
  \citenamefont {Zeyher},\ and\ \citenamefont {Birman}}]{brenig1972}%
  \BibitemOpen
  \bibfield  {author} {\bibinfo {author} {\bibfnamefont {Wilhelm}\ \bibnamefont
  {Brenig}}, \bibinfo {author} {\bibfnamefont {Roland}\ \bibnamefont {Zeyher}},
  \ and\ \bibinfo {author} {\bibfnamefont {Joseph~L.}\ \bibnamefont {Birman}},\
  }\bibfield  {title} {\enquote {\bibinfo {title} {{S}patial {D}ispersion
  {E}ffects in {R}esonant {P}olariton {S}cattering. {II}. {R}esonant
  {B}rillouin {S}cattering},}\ }\href {\doibase 10.1103/PhysRevB.6.4617}
  {\bibfield  {journal} {\bibinfo  {journal} {Phys. Rev. B}\ }\textbf {\bibinfo
  {volume} {6}},\ \bibinfo {pages} {4617--4622} (\bibinfo {year}
  {1972})}\BibitemShut {NoStop}%
\end{thebibliography}

%

\end{document}